\documentclass[printer]{aa}  
\usepackage{graphicx}
\usepackage{txfonts}
\usepackage{hyperref}
\usepackage{verbatim}
 \usepackage{longtable,lscape}
 \usepackage{hyperref}
 \usepackage{epsfig}
\usepackage{natbib}
\usepackage{xspace}

\usepackage{setspace}\usepackage{threeparttable}

\bibpunct{(}{)}{;}{a}{}{,} 

\newcommand{\gbm}{GBM\xspace}
\newcommand{\rhessi}{RHESSI\xspace}
\newcommand{\fermi}{\emph{Fermi}\xspace}

\begin{document}

\title{Quasi-Periodic Pulsations in Solar Flares:  \\
       new clues from the \fermi Gamma-Ray Burst Monitor}


   \author{ D.Gruber\inst{1},
   		P. Lachowicz\inst{2},
		E. Bissaldi\inst{3},
		M. S. Briggs\inst{4},
		V. Connaughton\inst{4},
		J. Greiner\inst{1},
		A. J. van der Horst\inst{4},\\
		G. Kanbach\inst{1},
		A. Rau\inst{1},
		P. N. Bhat\inst{4},
		R. Diehl\inst{1},
		A. von Kienlin\inst{1},
		R. M. Kippen\inst{5},
		C. A. Meegan\inst{6},
		W. S. Paciesas\inst{4},\\
		R. D. Preece\inst{4},
		C. Wilson-Hodge\inst{3}.
          }

   \institute{Max Planck Institute for Extraterrestrial Physics, 
   		Giessenbachstrasse, Postfach 1312, D-85748, Garching, Germany\\
		\email{dgruber@mpe.mpg.de}
		\and
           Green Cross Capital Pty Ltd, 495 Harris St, Ultimo,
           NSW 2007, Australia
           	\and
		Institute of Astro and Particle Physics, University Innsbruck,
		Technikerstrasse 25, 6176 Innsbruck, Austria
		\and
		University of Alabama in Huntsville, NSSTC, 
		320 Sparkman Drive, Huntsville, AL 35805, USA 
		\and
		 Los Alamos National Laboratory, 
		 P.O. Box 1663, Los Alamos, NM 87545, USA
		  \and
		 Universities Space Research Association, 
		 NSSTC, 320 Sparkman Drive, Huntsville, AL 35805, USA
		\and
		Space Science Office, VP62, NASA/Marshall Space Flight Center
		Huntsville, AL 35812, USA	
		  }


 
  \abstract
  {}
  {In the last four decades it has been observed that solar flares show quasi-periodic pulsations (QPPs) from the lowest, i.e. radio, to the highest, i.e. gamma-ray, part of the electromagnetic spectrum. To this day, it is still unclear which mechanism creates such QPPs. In this paper, we analyze four bright solar flares which show compelling signatures of quasi-periodic behavior and were observed with the Gamma-Ray Burst Monitor (\gbm) onboard the \fermi satellite. Because \gbm covers over 3 decades in energy (8~keV to 40~MeV) it can be a key instrument to understand the physical processes which drive solar flares.}
  {We tested for periodicity in the time series of the solar flares observed by \gbm by applying a classical periodogram analysis. However, contrary to previous authors, we did not detrend the raw light curve before creating the power spectral density spectrum (PSD). To assess the significance of the frequencies we made use of a method which is commonly applied for X-ray binaries and Seyfert galaxies. This technique takes into account the underlying continuum of the PSD which for all of these sources has a $P(f)\sim f^{-\alpha}$ dependence and is typically labeled red-noise.} 
  {We checked the reliability of this technique by applying it to a solar flare which was observed by the Reuven Ramaty High-Energy Solar Spectroscopic Imager (\rhessi) which contains, besides any potential periodicity from the Sun, a 4~s rotational period due to the rotation of the spacecraft around its axis. While we do not find an intrinsic solar quasi-periodic pulsation we do reproduce the instrumental periodicity. Moreover, with the method adopted here, we do not detect significant QPPs in the four bright solar flares observed by \gbm. We stress that for the purpose of such kind of analyses it is of uttermost importance to appropriately account for the red-noise component in the PSD of these astrophysical sources.}
  {}


  \titlerunning{About QPPs in solar flares: new clues from \fermi/GBM}
  \authorrunning{D. Gruber et al.}

  \maketitle
%

\section{Introduction}

Over the last 40 years quasi-periodic pulsations (QPP) in solar flares have been reported from observations across the electromagnetic spectrum, i.e. from radio waves up to the high energetic gamma-rays ranging from sub-second timescales up to several minutes  \citep[e.g.][]{parks69, ofsu06, ligan08, nakmel09, nak10}. While there seems to be an overwhelming amount of observational data, 
the underlying physical mechanism which could generate such QPPs remains still a mystery. 
Possible processes that are considered are modulation of electron dynamics by magnetohydrodynamic (MHD)
oscillations \citep{zait82}, periodic triggering of energy releases by MHD waves \citep{foull05, nakfoull06}, 
MHD flow overstabilities \citep{ofsu06} and oscillatory regimes of magnetic reconnection \citep{kliem00}.

In this paper we will present time series and periodogram analyses of four solar flares which show an intriguing quasi-periodic behavior in their light curves. All of these solar flares were observed by the \fermi Gamma-Ray Burst Monitor
\citep[GBM,][]{meegan09}. \gbm is one of the instruments onboard the
\fermi Gamma-Ray Space Telescope \citep{atwood09} launched on
June 11, 2008. Specifically designed for gamma-ray burst (GRB) studies, \gbm observes the whole unocculted sky with a total of 12 thallium-activated sodium iodide (NaI(Tl)) scintillation detectors covering the energy range from 8~keV to 1~MeV and two bismuth germanate scintillation detectors (BGO) sensitive to energies between 150~keV and 40~MeV \citep{meegan09}.
Therefore, \gbm offers superb capabilities for the analyses of not only GRBs but solar flares as well.

The analysis and interpretation of power spectral density (PSD) of solar flares is, in general, difficult. 
A variety of astrophysical sources (such as X-ray binaries, 
Seyfert galaxies \citep[e.g.][]{lawrence87, mark03} and GRBs \citep{ukwatta09, cenko10}),
show erratic, aperiodic brightness changes. Solar flares exhibit similar aperiodic variations with the general time profile being a sharp impulsive phase followed by a slower decay phase. Solar flares, together with many other astrophysical sources, thus have very steep power spectra in the low-frequency region.
This type of variability is known as red-noise \citep[e.g.][]{groth75, deeter82,vaughan05}. When determining the significance of possible periodicities in the PSD the red-noise has to be accounted for in order not to severely overestimate the significance of identified frequencies \citep{lachgup09}. In this paper we will account for the red-noise properties when performing a periodogram analysis.

This paper is organized as follows. In Sect.\ref{sec:rednoise} we briefly present the methodology of the time-series analysis. We provide an overview of the red-noise properties in astrophysical sources and demonstrate the importance of the red-noise when estimating significances. In Sect.\ref{sec:gbm} we will present light curves and periodograms of solar flares which were observed by \gbm. Finally, in Sect.\ref{sec:conc}, we will summarize and conclude.

\section{Analysis of data governed by red-noise}
\label{sec:rednoise}

As it was already pointed out by \citet{mandel69} and \citet{press78}, the human eye has a tendency to 
identify periodicities from purely random time series, i.e. where sinusoidal variations are 
not statistically real. According to  \citet{press78} the strongest eye-apparent period in (actually non-periodic)
data will be about one-third the length of the data sample. According to these authors ``three-cycle'' quasiperiods should be taken with a grain of salt.

Solar flares fall into the group of astrophysical sources where red-noise is important.
Red-noise has nothing to do
with measurement errors or systematics of the detectors, which are also called noise. Red-noise
is an \textit{intrinsic} property of the observed source and is due to erratic, aperiodic brightness changes. Contrary to white noise, which displays a flat spectrum in a PSD, i.e. is power independent of frequency,
red-noise is characterized by a power law of the form of $P=Nf^{-\alpha}$.
As a first order approximation, red-noise is the realization of a linear stochastic and weakly non-stationary process.
This red-noise component makes the interpretation of the significance of a peak in the PSD more complex.

One way to estimate the significance of induced frequencies on top of an underlying 
red-noise continuum in a PSD, was presented by \citet{vaughan05}.

In short, \citeauthor{vaughan05} suggests to calculate the periodogram normalized 
so that the units of power are $\rm{(rms/mean)}^2 \rm{Hz}^{-1}$ \citep[e.g.][]{schuster1898, press89}. 
Then, the periodogram is converted to log-space in both frequency and power. For such a log-periodogram
one can than clearly identify the power-law component in the low-frequency range and a ``cutoff''
where white noise or an additional noise component takes over \citep[e.g.][]{ukwatta09}.
One can easily determine the power-law parameters by fitting a linear function to the low-frequency
periodogram bins using the least-squares method. 
In this paper the method of  \citet{vaughan05} was slightly modified in that we use a broken power law (BPL) to fit the PSD instead of a single power law.


\subsection{Red-noise simulation}
It is common practice \citep{inglis08, nak10} to suppress the low-frequency 
component by de-trending the light curves of solar flares. This can be achieved by smoothing the light curve with a moving average or by
applying a Gaussian filter and perform 
the periodogram analysis on the residual emission, i.e. the smoothed version is subtracted from the original data-set.
This can give rise to misleading results as we will show in the following.

\begin{figure}
\centering
\includegraphics[angle=0,width=0.5\textwidth]{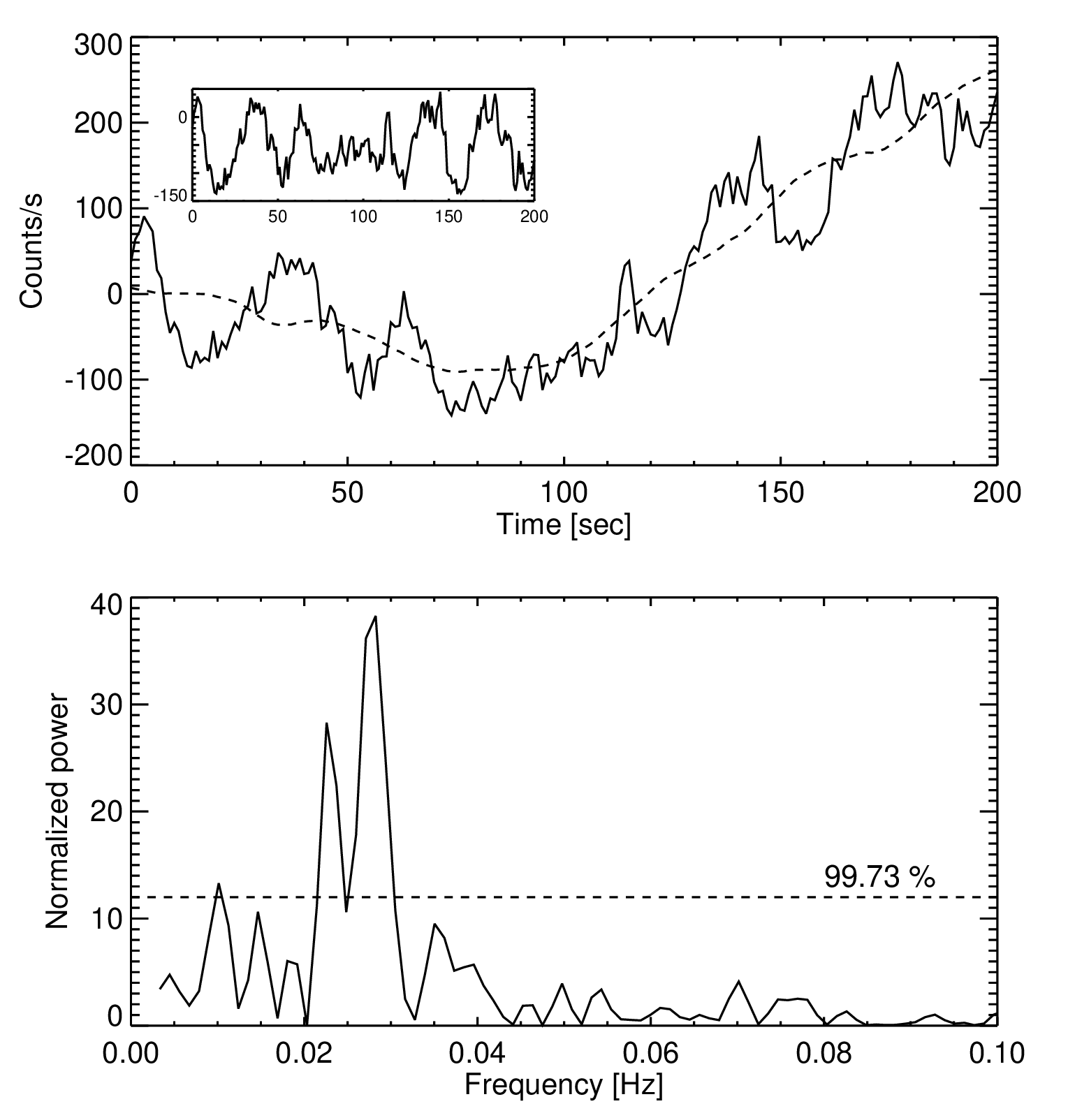}
\includegraphics[angle=0,width=0.5\textwidth]{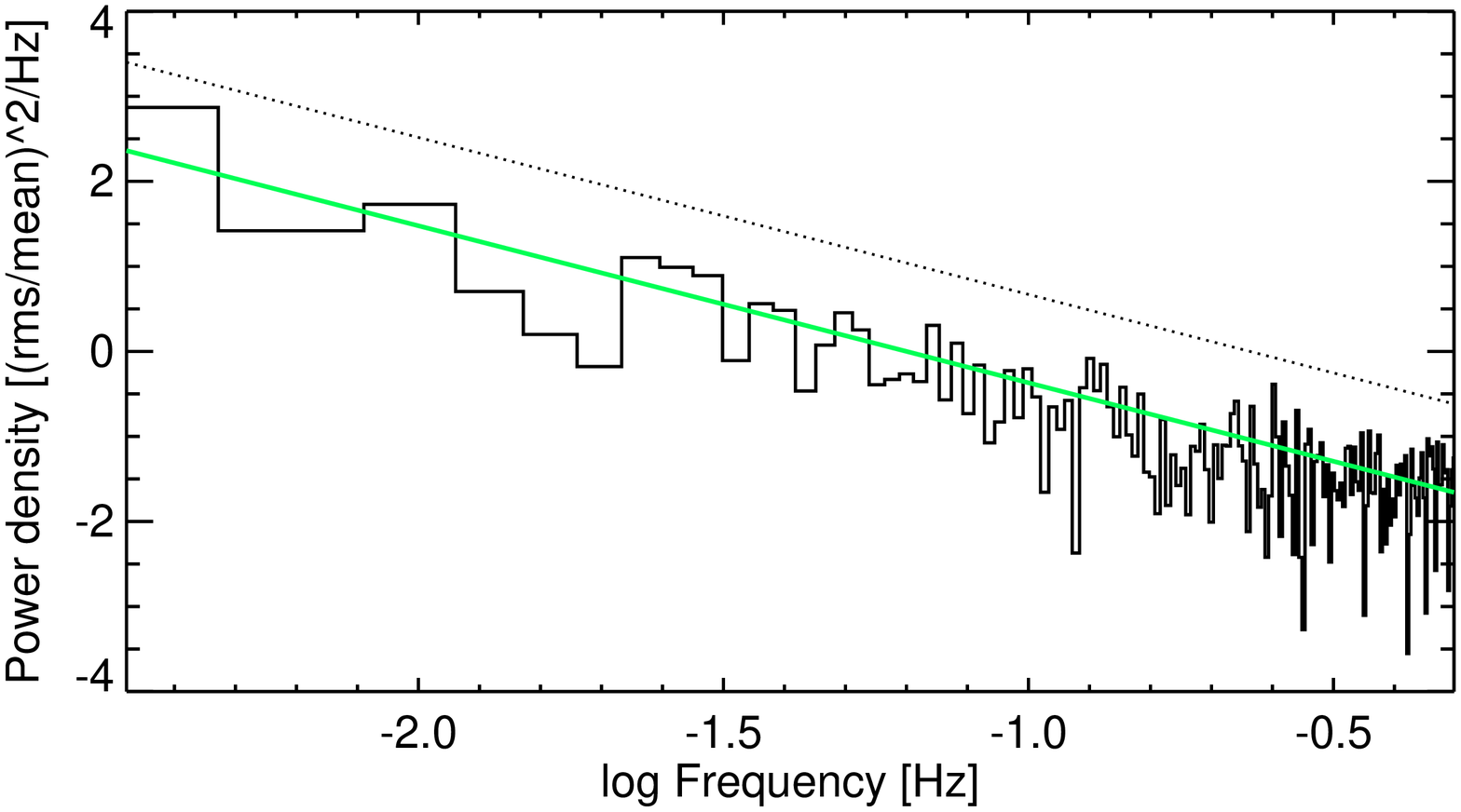}
\caption{\textit{Upper panel:} Synthetic red-noise time series with a power law index of $\alpha=-1.8$. Simple (boxcar) moving average is overplotted (dashed line). The inset presents the residual signal. \textit{Middle panel:} Lomb-Scargle Periodogram of the residual light curve. The $3\sigma$ CL has been denoted by a dashed line. \textit{Lower panel:} PSD with the best fit power law (solid green line) and the $3\sigma$ significance level (dotted line). \label{rednoise}}
\end{figure}

By randomizing the phase as well as the amplitudes, \citet{tiko95} introduced an algorithm
to generate a purely random time series which shows a $1/f^{\alpha}$
dependence in the PSD. We created a time series 
consisting of 300 data points, evenly spaced by 1~s with a $f^{-1.8}$ periodogram shape. 
From this light curve, we subtracted
a simple moving average of 50~s (see Fig. \ref{rednoise}).
We then calculated the PSD (first described by \citet{lomb76} 
and \citet{scargle82}) and then 
refined by \citealt{press89}) on the residuals. The result is striking. Although we 
started with a purely random, red-noise dominated time-series 
we obtain a PSD with three 
frequencies whose power exceeds the $3\sigma$ confidence limit (calculated according to \citealt{scargle82}).
This approach in signal processing clearly returns false-positive frequencies with periods of $P\approx 33$~s, $P\approx 55$~s
and $P\approx100$~s, respectively. 
When the periodogram is calculated over the original light curve, using the
\citet{vaughan05} approach, i.e. with a single power law fit to the data,
the spectral peaks remain below the $3\sigma$ threshold.

Therefore, by this example, we strongly discourage subtracting smoothed versions of raw light curves when 
looking for intrinsic frequencies in the red-noise dominated PSD.

\subsection{Detour: QPPs in GRBs}
The same procedure of light curve reprocessing with a subsequent PSD analysis has also been applied for GRBs. These events are the most luminous flashes of $\gamma$-rays known to mankind, believed to originate from highly relativistic outflows from a compact source with Lorentz factors $\Gamma > 100$. In 2009, the Swift-\textit{BAT} satellite \citep{gehrels04} observed  GRB~090709A \citep{morris09}. Soon after the detection, \citet{markwardt09} claimed to have found a very unusual behavior, not observed in any other GRB so far, namely a QPP like behavior with a periodicity of $P\approx 8.1$~s at the 12~$\sigma$ level of significance. This QPP was subsequently confirmed and found to be in phase in the data of the Anticoincidence System (ACS) of the spectrometer SPI on board the INTEGRAL satellite \citep{gotz09}, the Konus-WIND instrument \citep{gol09} and the Suzaku Wide-band All-sky Monitor (WAM) \citep{ohno09}. These latter instruments operate in the energy ranges 80~keV-10~MeV, 20~keV-1150~keV and 50~keV-5~MeV, respectively. Swift-\textit{BAT}, on the other hand, is sensitive in the energy range 15~keV-150~keV. 

However, soon thereafter, \citet{cenko10} showed that the interpretation of this QPP strongly depends on the assumption of the underlying continuum. If, in fact, it is accounted for, the significance of the claimed periodicity drops below the 3~$\sigma$ confidence limit. This analysis was independently repeated by \citet{Iwakiri10} and \citet{deLuca10} who also took into account the red-noise component in the PSD, and only find a marginally significant periodicity at the $3\sigma$ confidence limit. 

In conclusion to this detour, we emphasize once more the importance to account for the red-noise component in the PSD. Additionally, we draw attention to the fact that a potential quasi-periodic signal is not necessarily significant even if it is identified in several instruments with different (but overlapping) energy ranges and observed to be in phase across these bands.

\subsection{Method testing}
\begin{figure}
\centering
\includegraphics[angle=0,width=0.42\textwidth]{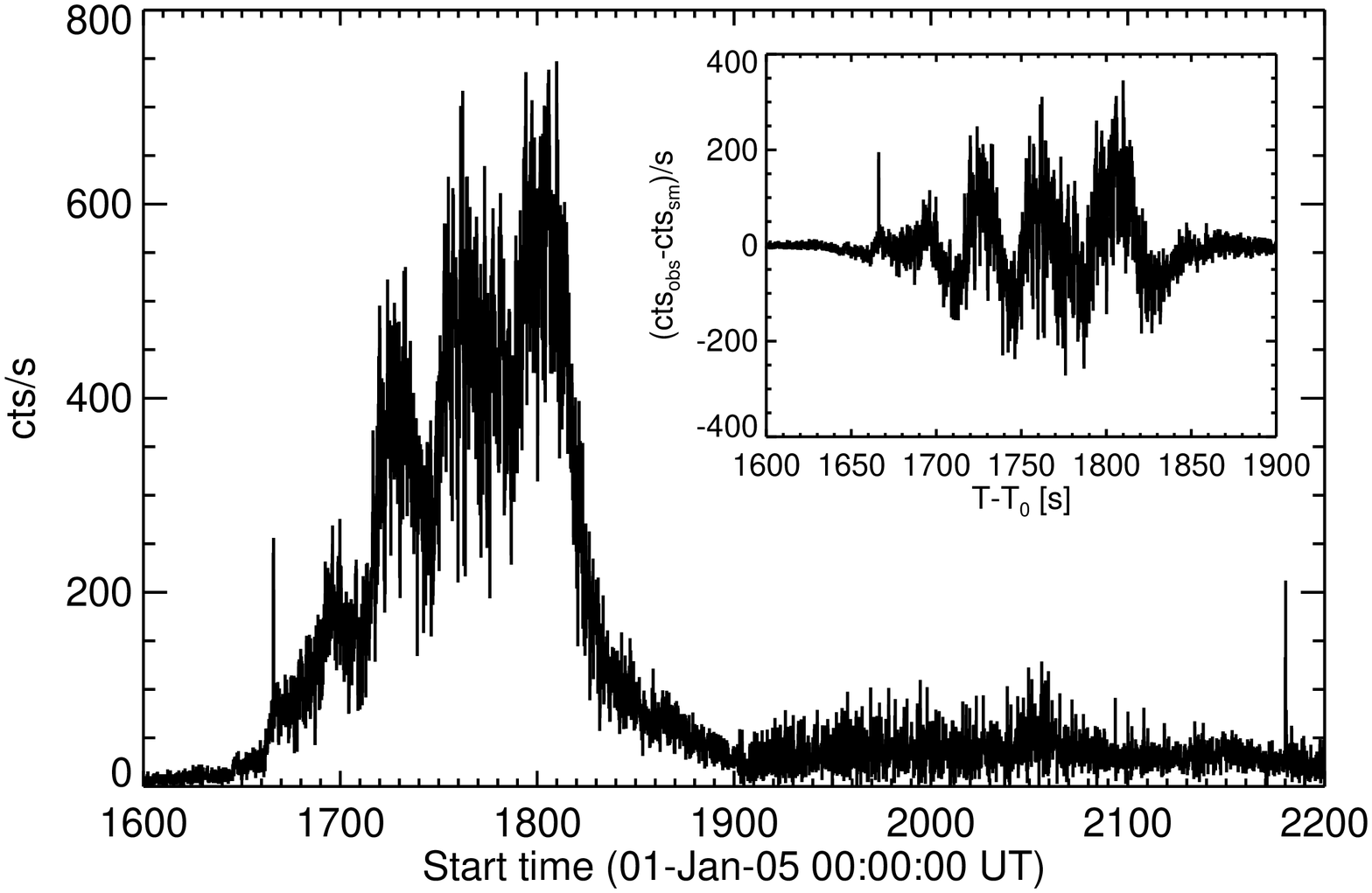}
\includegraphics[angle=0,width=0.42\textwidth]{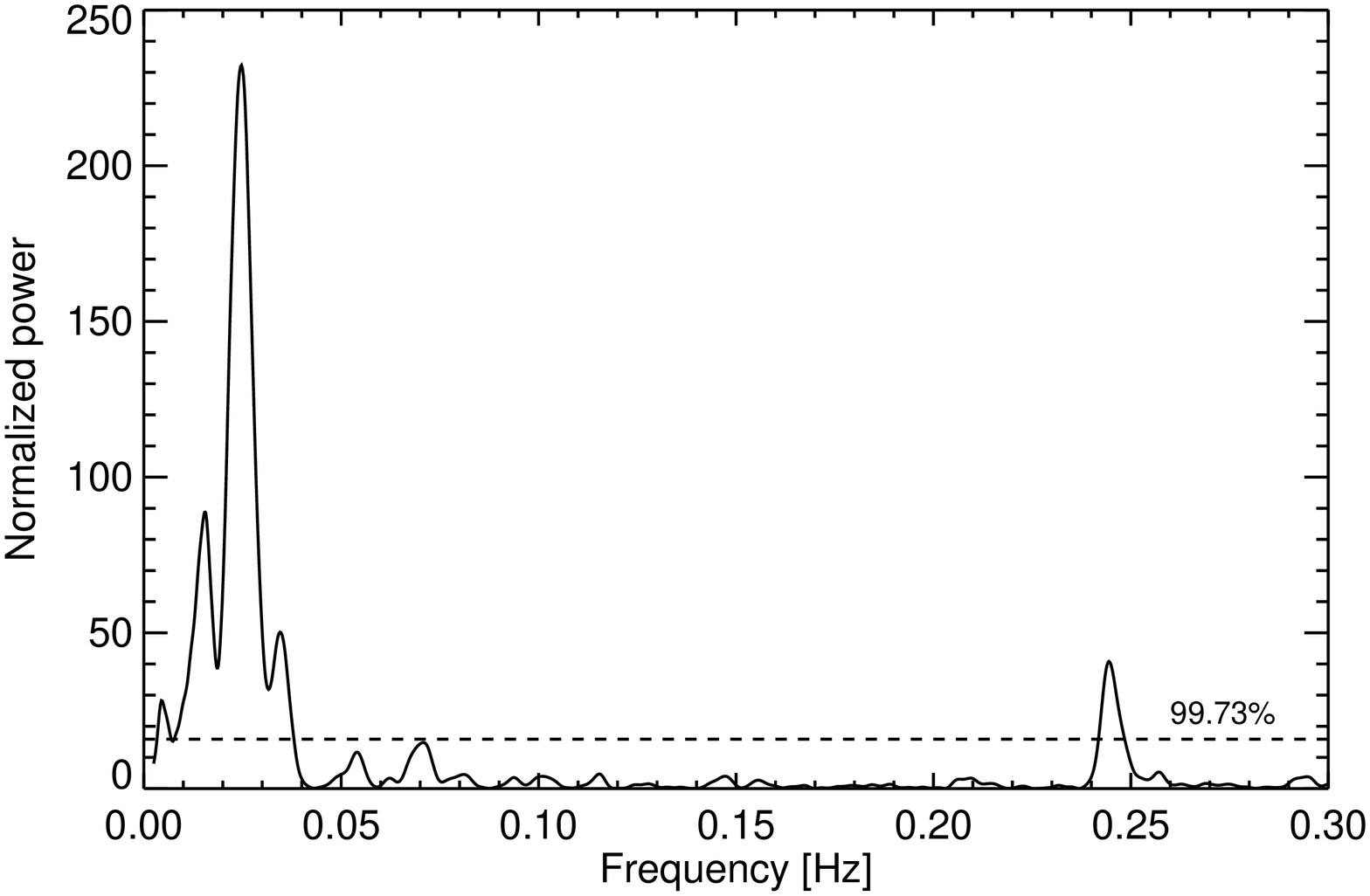}
\includegraphics[angle=0,width=0.42\textwidth]{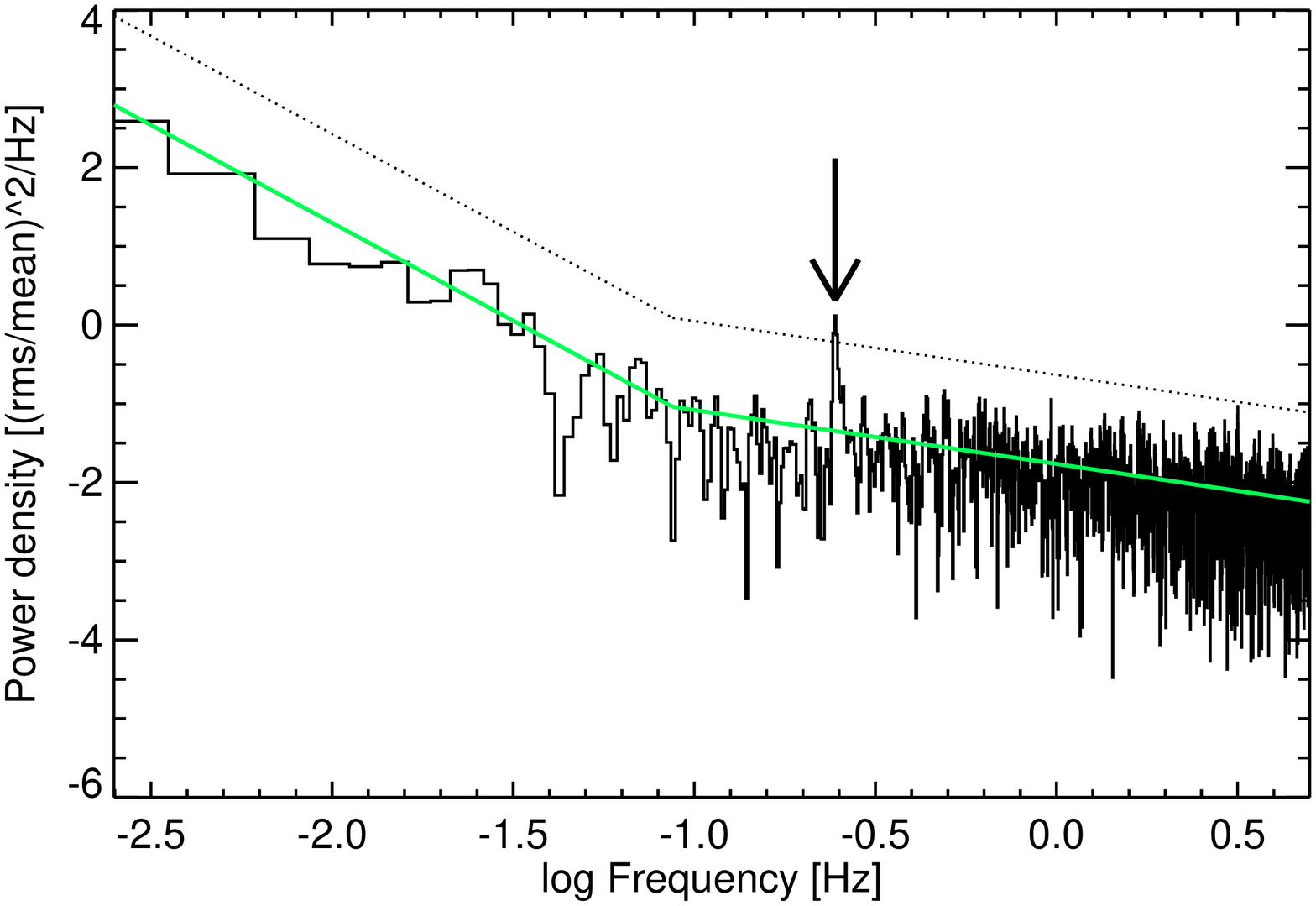}
\caption{\textit{Upper panel:} Summed and background subtracted light curve of the solar flare observed on January 1st, 2005 by \rhessi. The inset shows the residual emission after subtracting a simple (boxcar) moving average. \textit{Middle panel:} Periodogram analysis performed on the residual emission.
\textit{Lower panel:} PSD with best fit broken power-law (solid green line) and the $3\sigma$ significance level (dotted line) indicated. The arrow points to the significant frequency at $f \approx 0.244$~Hz.}
\label{fig:rhessi_050101_sfl}
\end{figure}

The Reuven Ramaty High-Energy Solar Spectroscopic Imager (\rhessi, \citealt{rhessi}) rotates around its spin axis which is always pointed towards the Sun. The period of this rotation is $P\approx 4$~s and a PSD analysis of the \rhessi light curves is expected to display this instrumental signal. We applied the \citet{vaughan05} test to a solar flare which has been observed by \rhessi on January 1st, 2005 and where QPPs have been reported \citep{nak10}. This solar flare peaked at 00:31 UT at a GOES level X1.7, from the NOAA active region 10715 located on disk at N03E47. We used RHESSI data in the energy range from 50~keV-100~keV 
with a fine time resolution of 0.1~s (see upper panel of Fig.\ref{fig:rhessi_050101_sfl}) and performed two periodogram analyses on this light curve in the range between 1660~s and 1820~s. The first periodogram analysis was performed using the classical approach introduced by \citet{lomb76} and \citet{scargle82}. Similarly to what has been commonly done in the past \citep[e.g.][]{inglis09} a periodogram has been calculated on the residual emission after a simple moving average of 50~s has been subtracted from the raw data (see middle panel of Fig.\ref{fig:rhessi_050101_sfl}). With this method, several peaks are found above the 3~$\sigma$ threshold. The peak with the highest value of normalized power is located at $f\sim0.025$~Hz, corresponding to the periodicity reported by \citet{nak10}. Another peak which is worth mentioning is located at $f\sim0.24$~Hz which is the expected rotational frequency of \rhessi around its spin axis.

As we will show in the following, the significance of the peak at $f\sim0.025$~Hz is highly overestimated by the latter method. We show a PSD which was calculated using the raw and undetrended light curve applying the technique by \citet{vaughan05} in the bottom panel of Fig.\ref{fig:rhessi_050101_sfl}. Analogically to the \citeauthor{lomb76} and \citeauthor{scargle82} periodogram analysis, we found a significant spectral feature at $\approx 0.24$~Hz which is the expected rotation period of the \rhessi instrument. However, this PSD is lacking any other frequency above the 3~$\sigma$ confidence limit. The discrepancy between the two methods is easily explained. Firstly, the whole raw light curve was used without artificially detrending it beforehand. Secondly, the method by \citeauthor{lomb76} and \citeauthor{scargle82} assumes a white noise continuum and does not take into consideration the red noise component of the solar flare. However, the latter is taken into account by the method of \citeauthor{vaughan05}. In conclusion, we could not confirm the reported QPP in this solar flare. We are confident that this latter method can be used for further analysis since we believe that it is more adequate for sources which are dominated by red-noise and the inherent rotational frequency of the instrument is found to be significant.

\begin{figure}
\centering
\includegraphics[angle=0,width=0.5\textwidth]{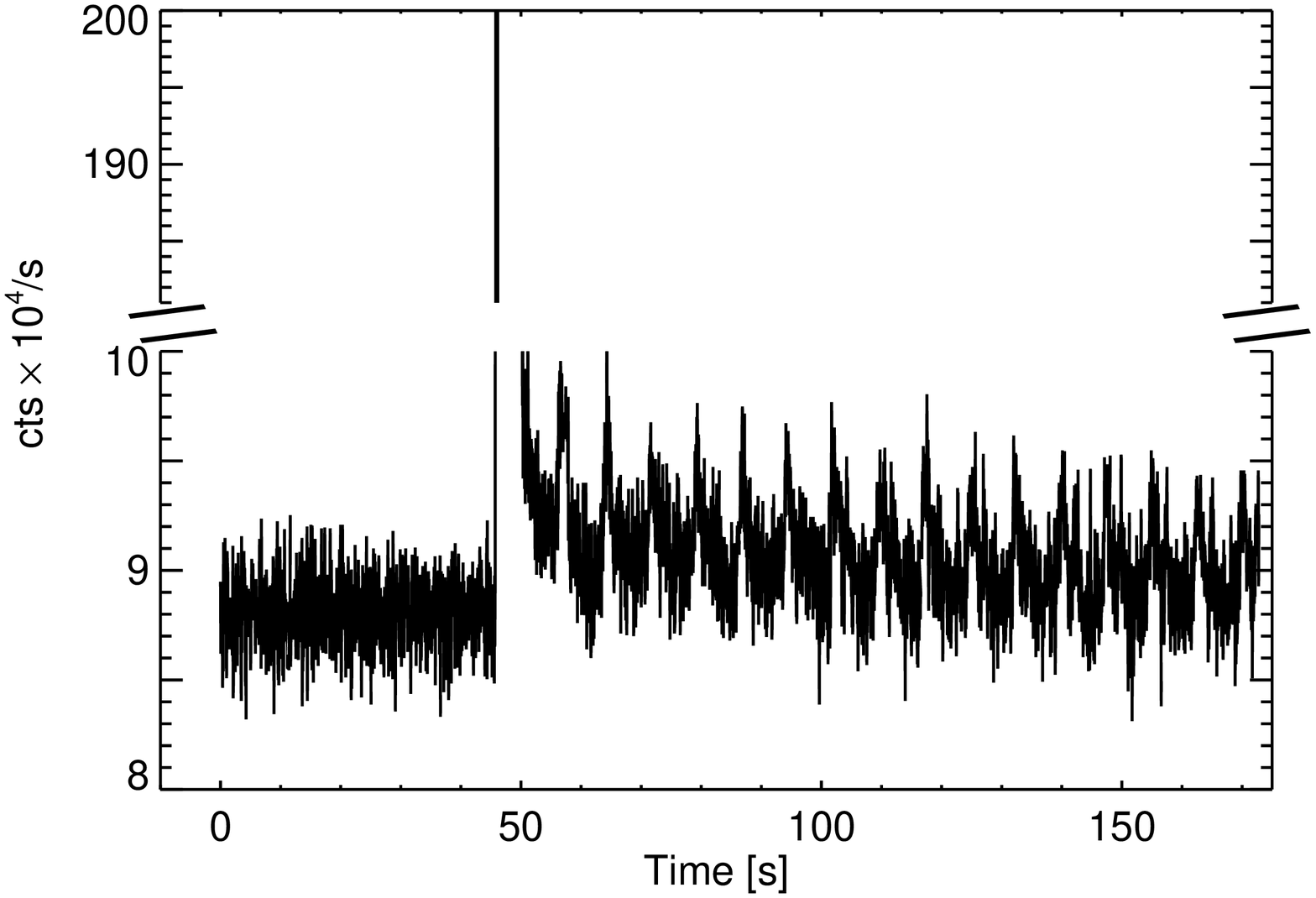}
\includegraphics[angle=0,width=0.5\textwidth]{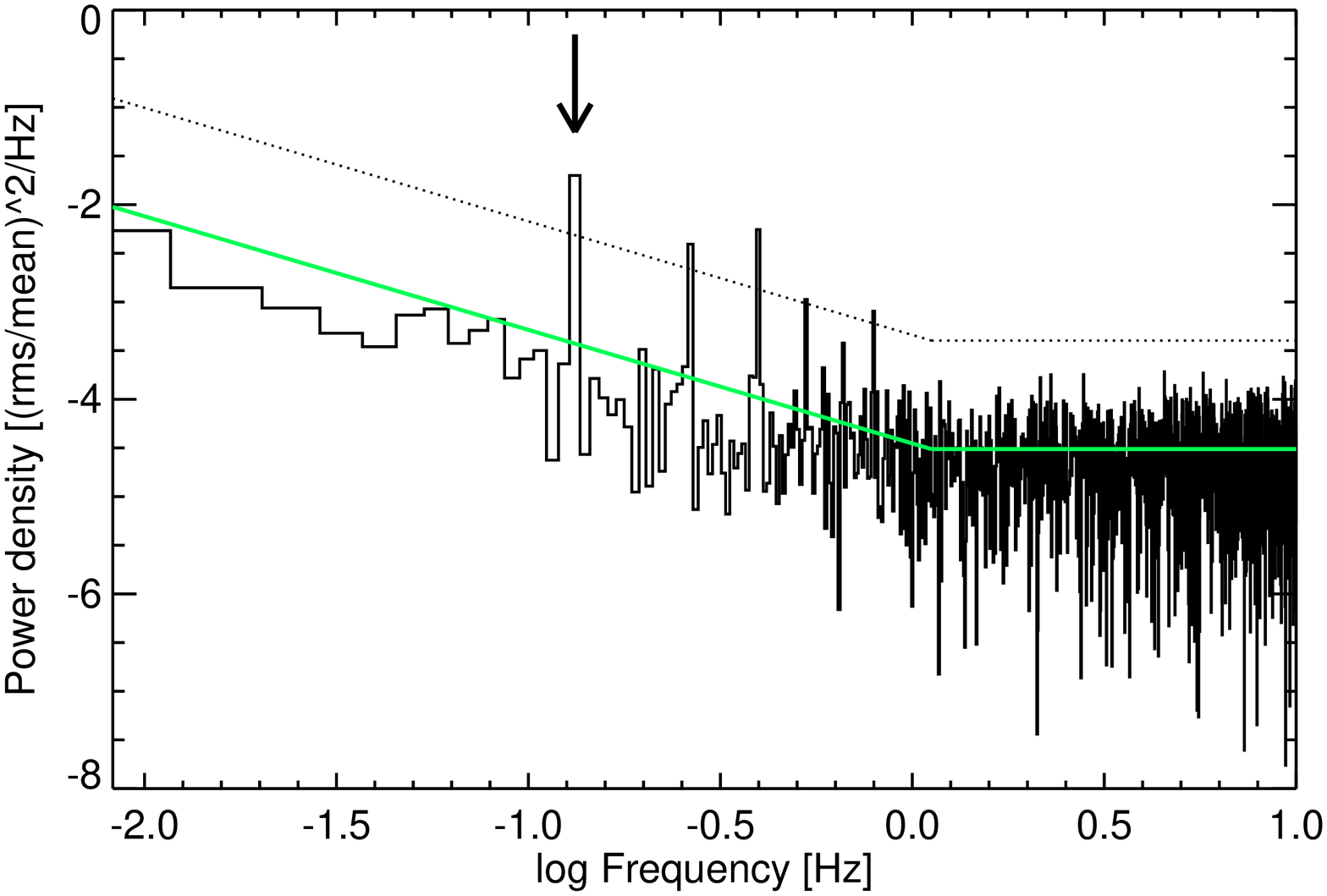}
\caption{\textit{Upper panel:} Light curve of the giant flare of SGR 1806-20 observed by SPI-ACS. \textit{Lower panel:} PSD of the SGR 1806-20 light curve. Best fit broken power-law to the PSD continuum (solid green line) and the $3\sigma$ significance level (dotted line) are indicated. The arrow indicates the recovered rotation period of $P\approx 7.56$~s. Other peaks above the $3 \sigma$ levels are harmonics.}
\label{fig:sgr}
\end{figure}

An additional check was performed using SPI-ACS data of the outburst of the soft gamma repeater SGR~1806-20 observed on December 27th, 2004 in the energy range from 80~keV to 8~MeV \citep{mere05}. SGR~1806-20 is known to have a rotational period of 7.56~s and the methodology adopted here should be able to recover this periodicity. We removed the very bright initial pulse and focused on the emission from $\approx 50$~s to 175~s (see Fig.\ref{fig:sgr}). We unambiguously recover the main pulsation period ($P\approx 7.56$~s) together with the first, second, third and fifth harmonic.

We conclude that the here applied methodology is appropriate and reliable for the further analysis.


\section{Solar flares observed by \gbm}
\label{sec:gbm}

\subsection{Solar flare on February 24, 2011 at 07:29:20.71 UT}
\begin{figure}
\centering
\includegraphics[angle=0,width=0.5\textwidth]{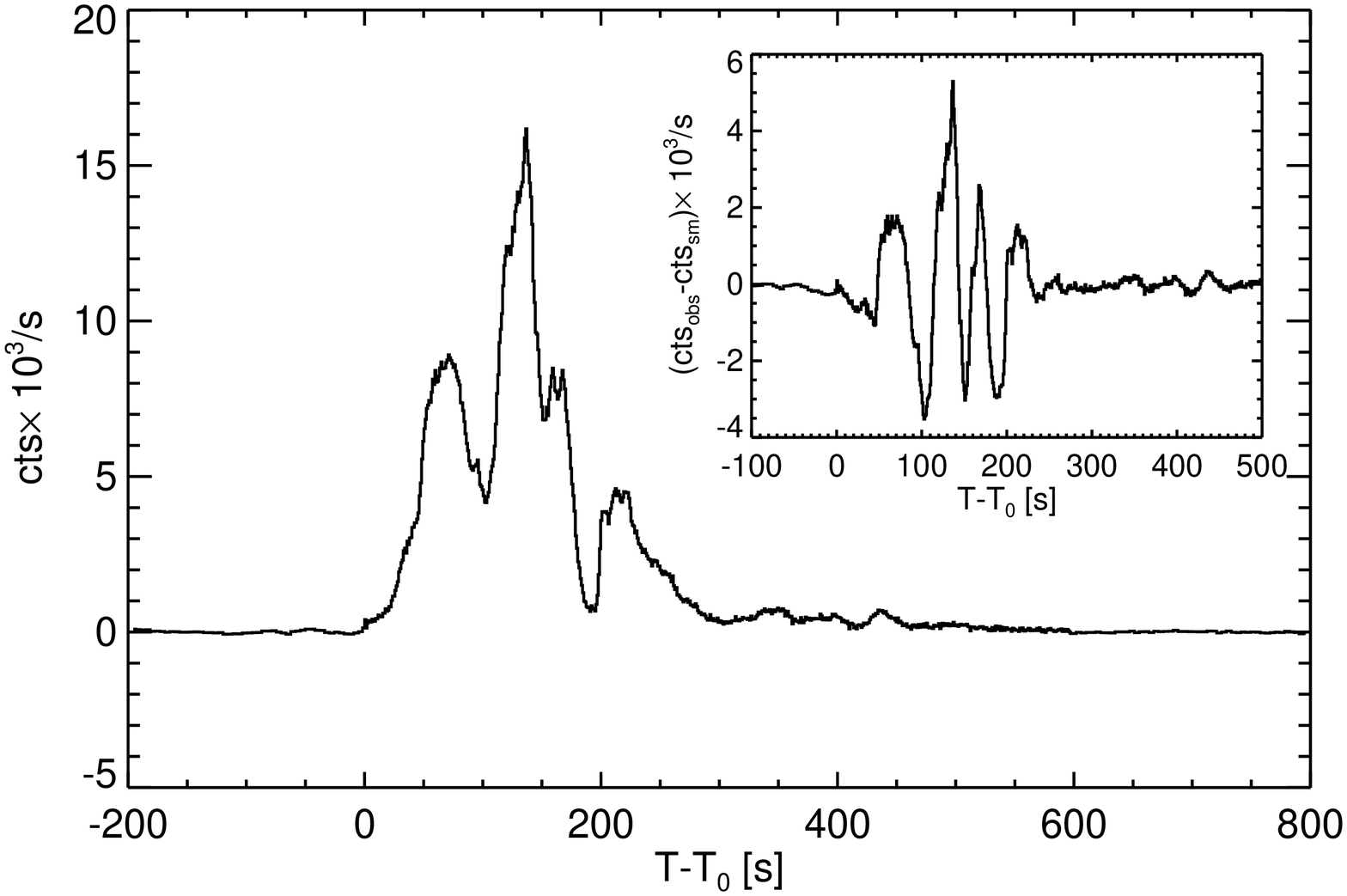}
\includegraphics[angle=0,width=0.5\textwidth]{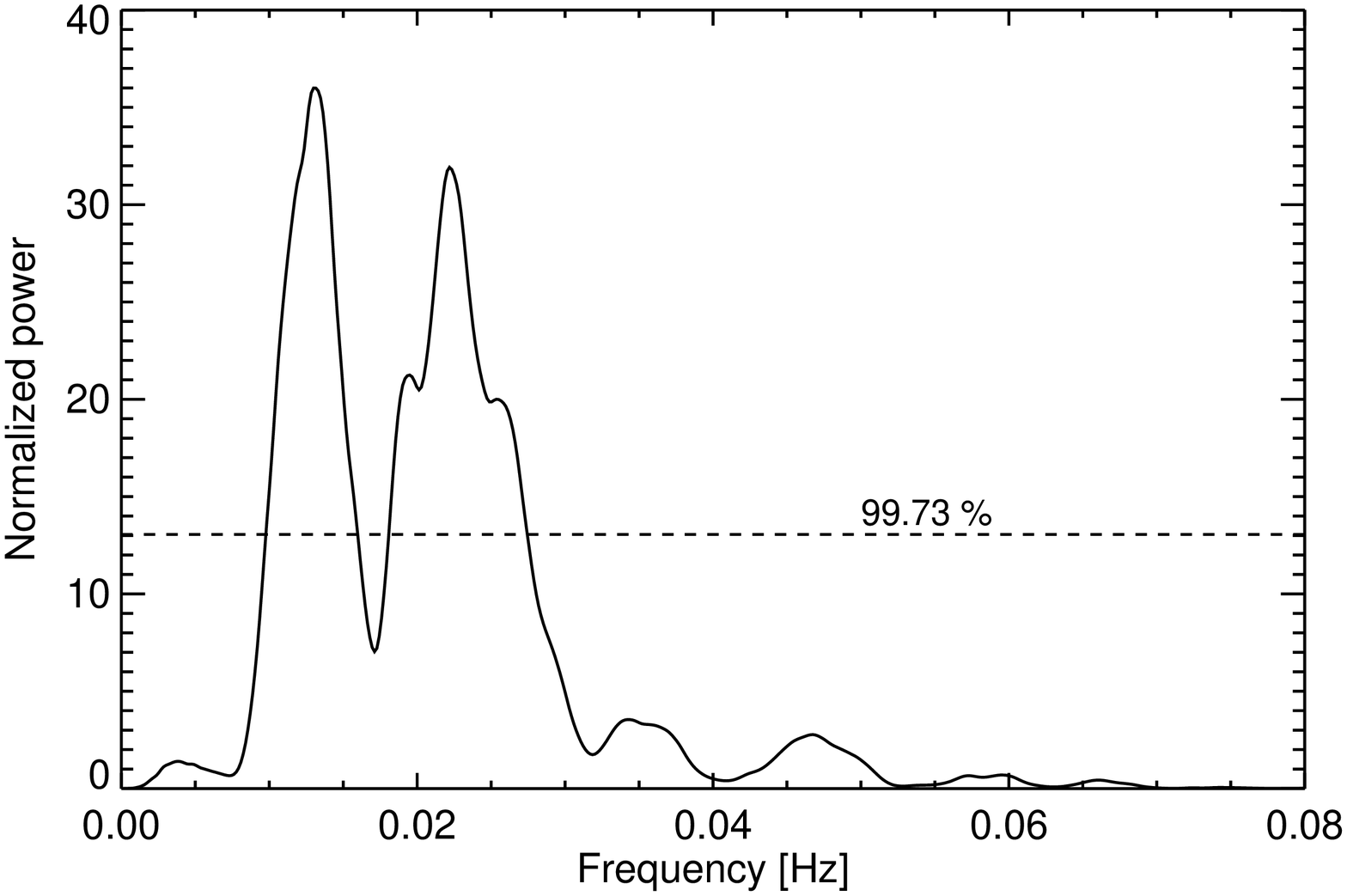}
\includegraphics[angle=0,width=0.5\textwidth]{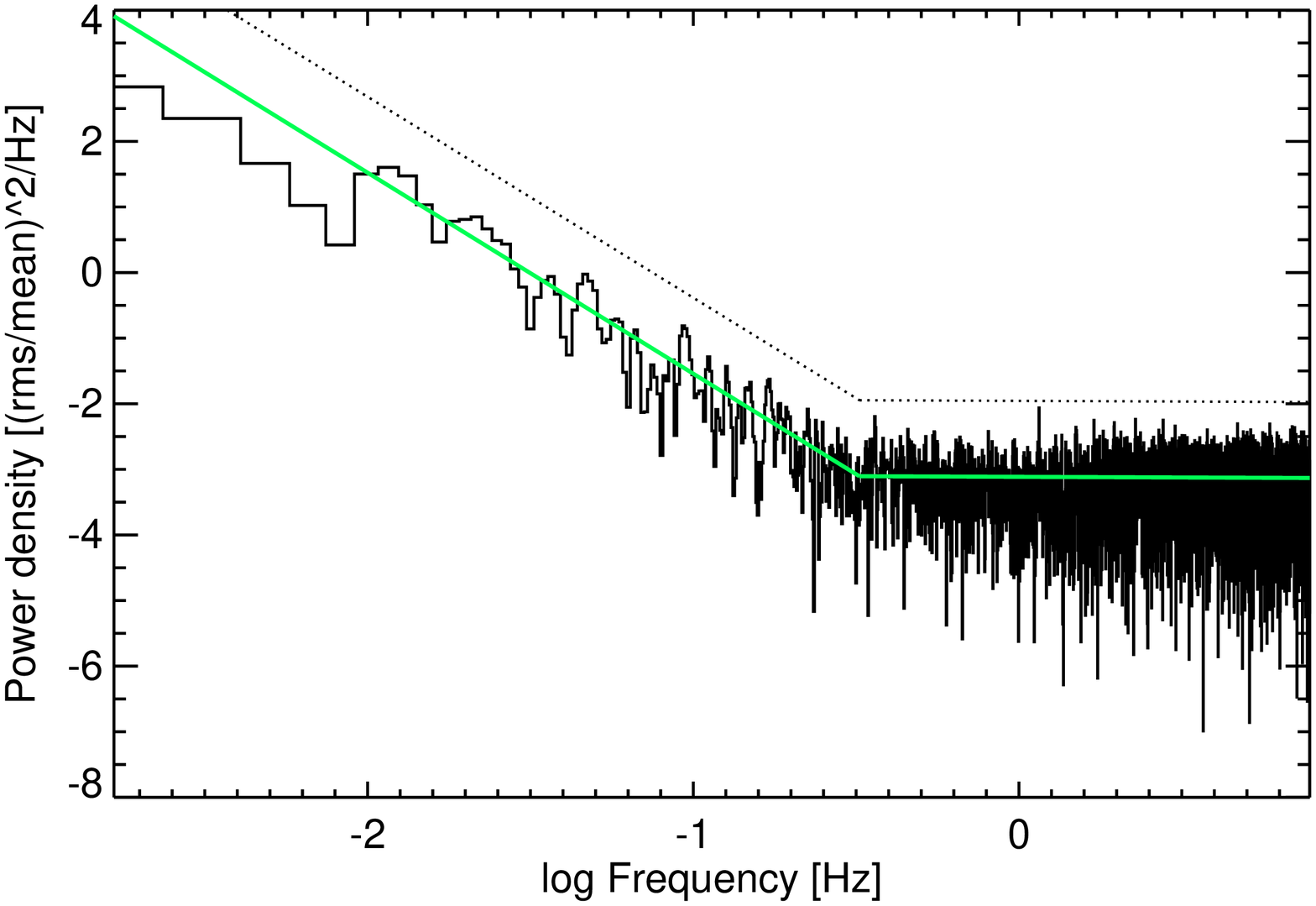}
\caption{\textit{Upper panel}: Summed and background subtracted light curve of the solar flare observed on February 24, 2011 by \gbm. The inset shows the residual emission after subtracting a simple (boxcar) moving average. \textit{Middle panel:} Periodogram analysis performed on the residual emission. Two peaks are above the $3\sigma$ confidence limit (dashed line).  \textit{Lower panel:} PSD of the same solar flare. Best fit broken power-law to the PSD continuum (solid green line) and the $3\sigma$ significance level (dotted line) are indicated. }
\label{fig:110224312_sfl}
\end{figure}

For the purpose of its analysis we use CSPEC and CTIME data of detectors NaI~3, NaI~4 and NaI~5 with a time-resolution of 1.024~s (4.096~s pre-trigger) and 0.064~s (0.256~s pre-trigger), respectively. In the energy range from 50~keV to 1~MeV this solar flare lasted for about 500~s. The light curve (Fig.\ref{fig:110224312_sfl}) consists of several peaks and a compellingly looking quasi-periodic behavior lasting to $\approx 500$~s. After de-trending the raw light curve with a simple moving average (50~s) the QPP pattern becomes more visible (see inset of Fig.\ref{fig:110224312_sfl}). Applying a standard periodogram analysis \citep{lomb76, scargle82} on the \textit{detrended} light curve several peaks are above the $3\sigma$ confidence limit as can be seen in the middle panel of Fig.\ref{fig:110224312_sfl}. 

However, applying the previously introduced method \citep{vaughan05} on the original light curve, thus taking into account the red-noise properties of the source we did not find any significant QPP during the solar flare life-time, as is pointed out in the lower panel of Fig.\ref{fig:110224312_sfl}.  We performed this analysis making use of the CTIME data with a time resolution of 64~ms. The PSD has been calculated for the signal spanning from ($T_0-10$)~s to ($T_0+600$~s), where $T_0$ denotes the time of trigger.

\subsection{Solar flare on June 12, 2010 at 00:55:05 UT}
\begin{figure}
\centering
\includegraphics[angle=0,width=0.5\textwidth]{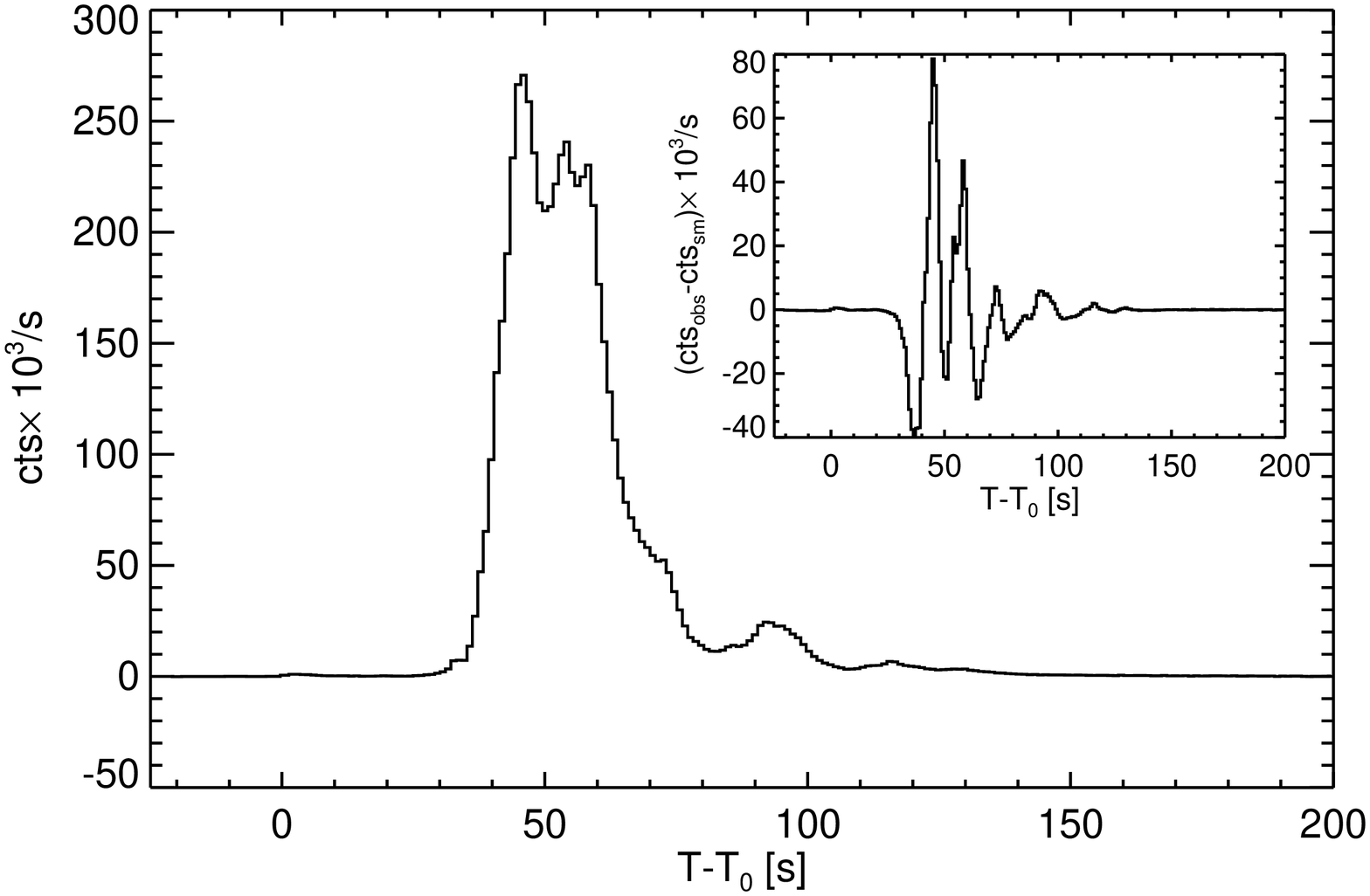}
\includegraphics[angle=0,width=0.5\textwidth]{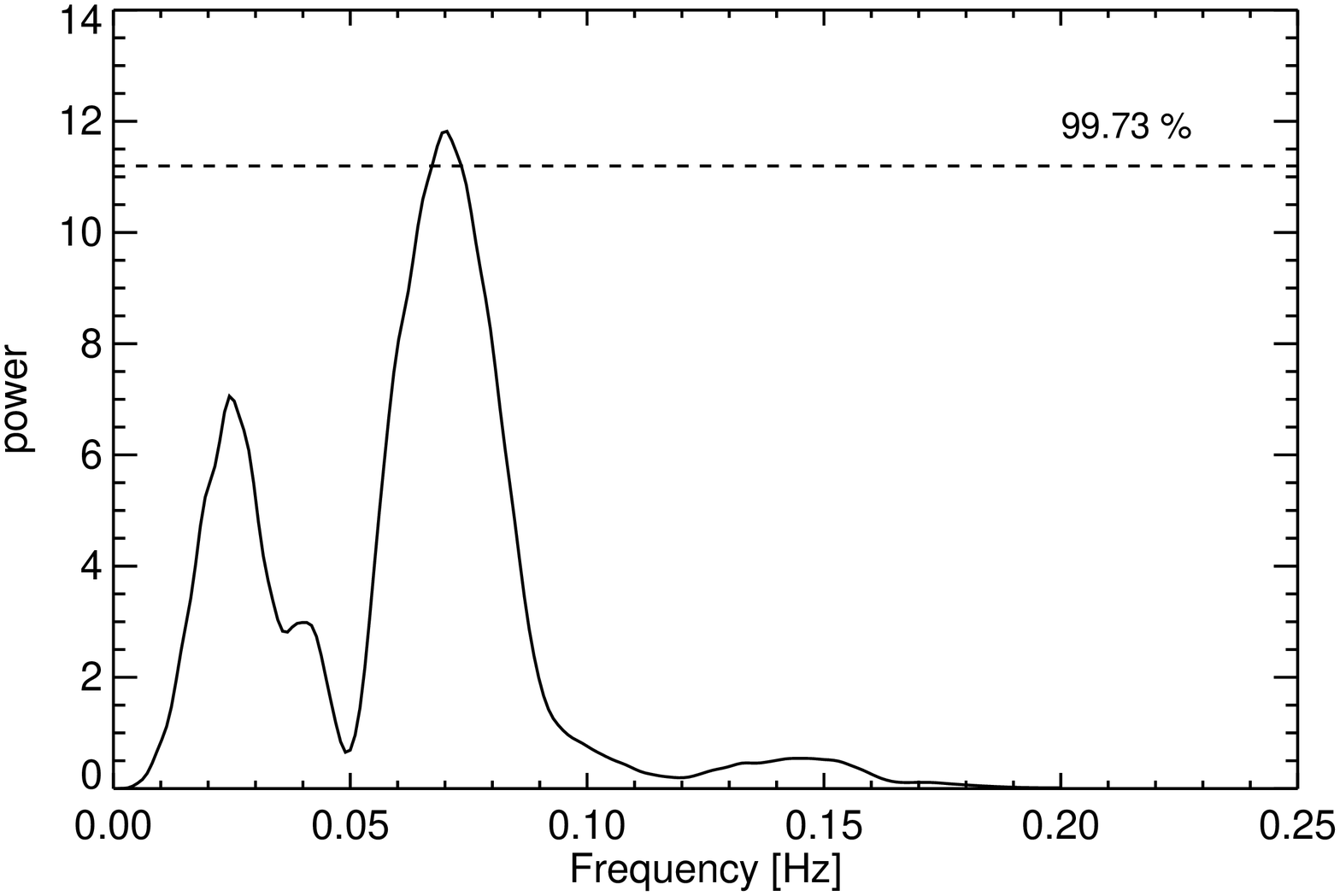}
\includegraphics[angle=0,width=0.5\textwidth]{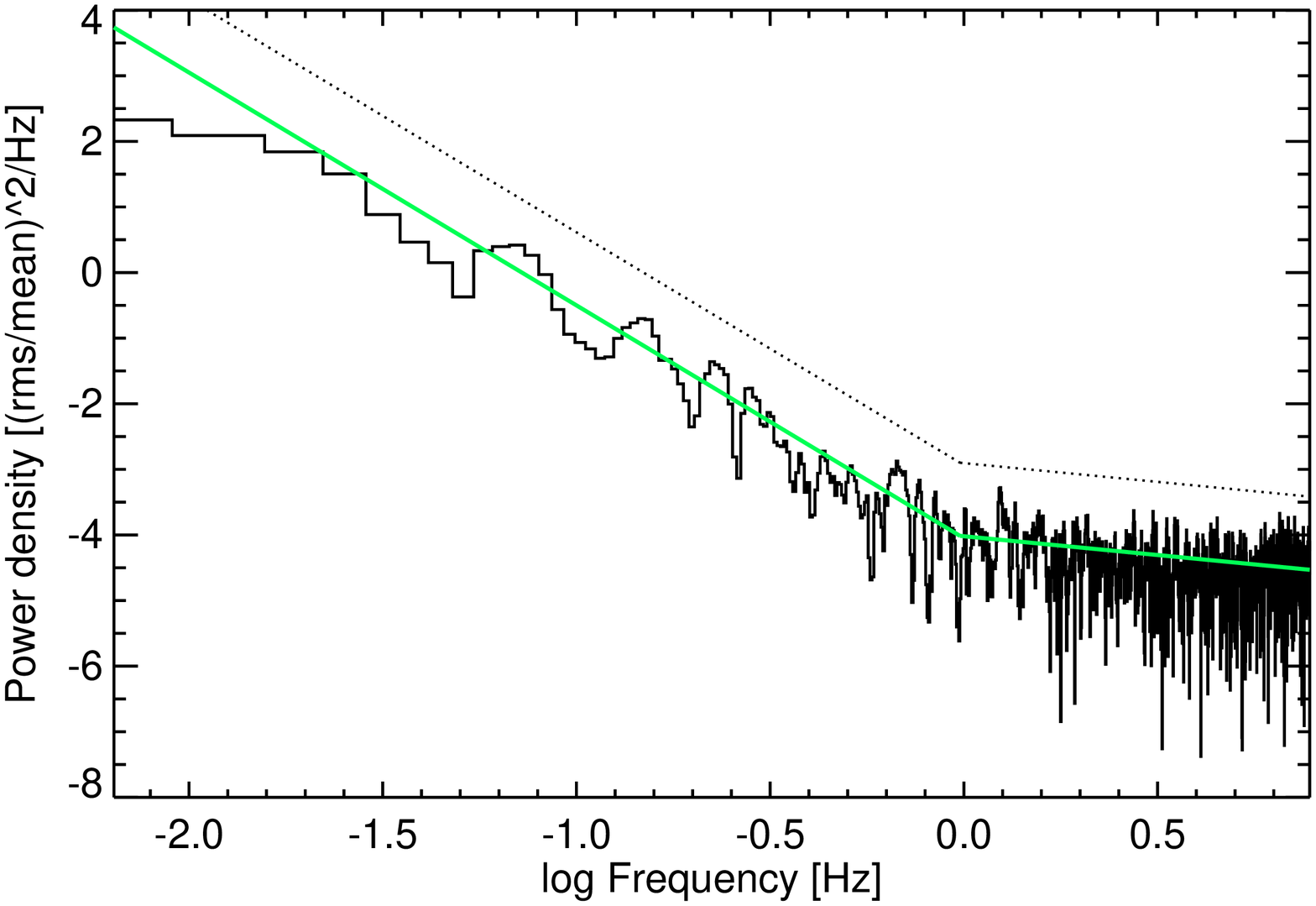}
\caption{\textit{Upper panel:} Same as Fig.\ref{fig:110224312_sfl} for the solar flare observed on June 12, 2010. \textit{Middle panel:} Same as Fig.\ref{fig:110224312_sfl}. \textit{Lower panel:} Same as Fig.\ref{fig:110224312_sfl}.}
\label{fig:100612038_sfl}
\end{figure}

For the analysis we use again CSPEC and CTIME data of detectors NaI 0 through NaI 5.
The count rate is not increasing significantly for the first 30~s in the 50~keV to 1~MeV  energy range. After this time there is a sharp 
increase in the flare brightness which then decays again very rapidly after 60~s. The whole duration of 
the solar flare in this energy range is approximately 120~s. 
Overlaid on top of the observed light curve, again one can identify a compellingly looking QPP behavior with a period of $\sim 15$~s (see upper panel of Fig.\ref{fig:100612038_sfl}). This periodicity is apparently significant when applying a standard Lomb-Scargle periodogram (see middle panel of Fig.\ref{fig:100612038_sfl}) on the detrended light curve. The smoothing length was 15~s.

However, the PSD, computed for CTIME data with a time resolution of 64~ms in the interval ($T_0-25$)~s to ($T_0+150$)~s and according to \citet{vaughan05}, does not show a significant periodicity.



\subsection{Solar flare on March 15, 2011 at 00:21:15.69 UT}
\label{sec:sfl4}
\begin{figure}
\centering
\includegraphics[angle=0,width=0.5\textwidth]{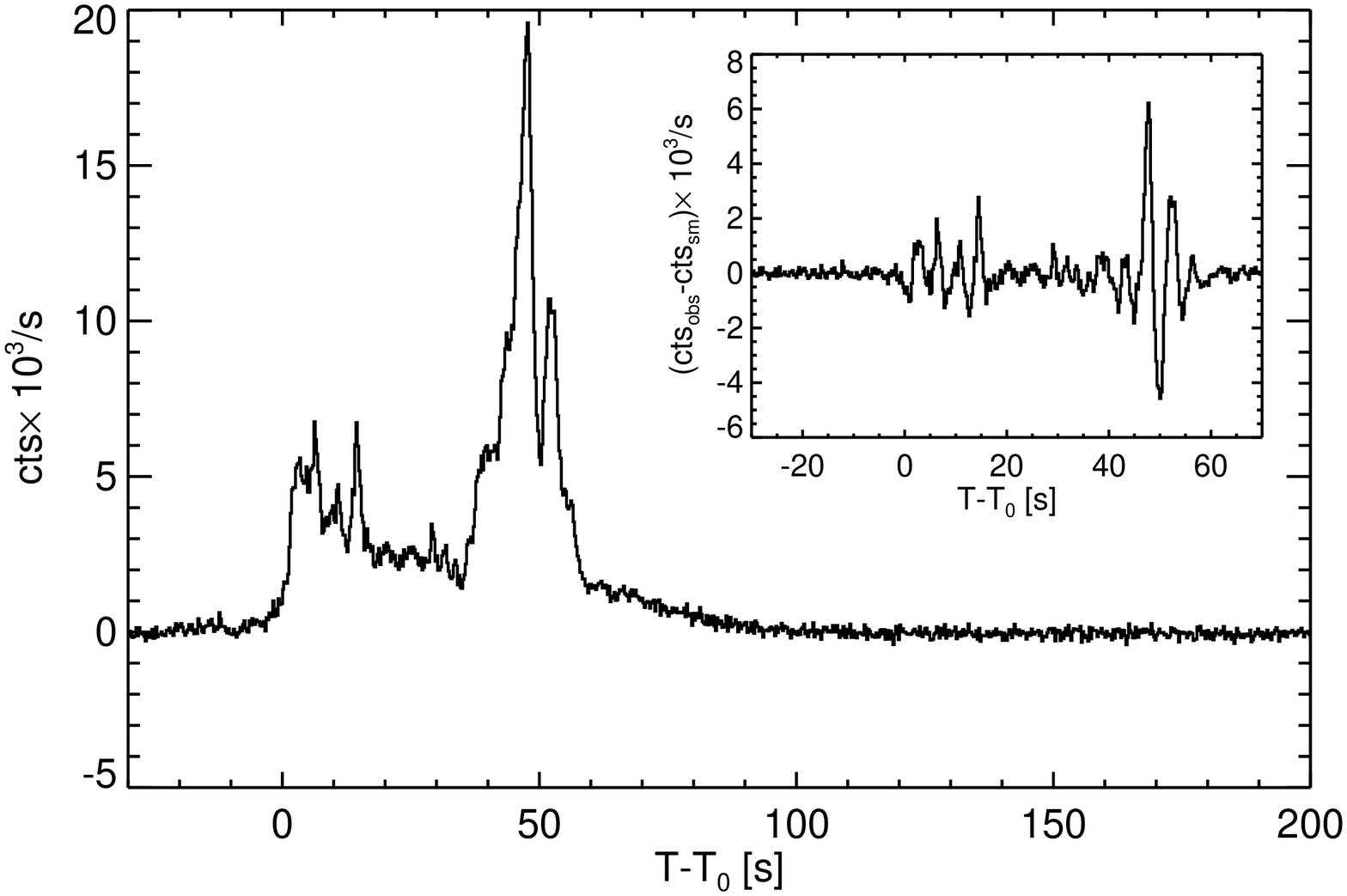}
\includegraphics[angle=0,width=0.5\textwidth]{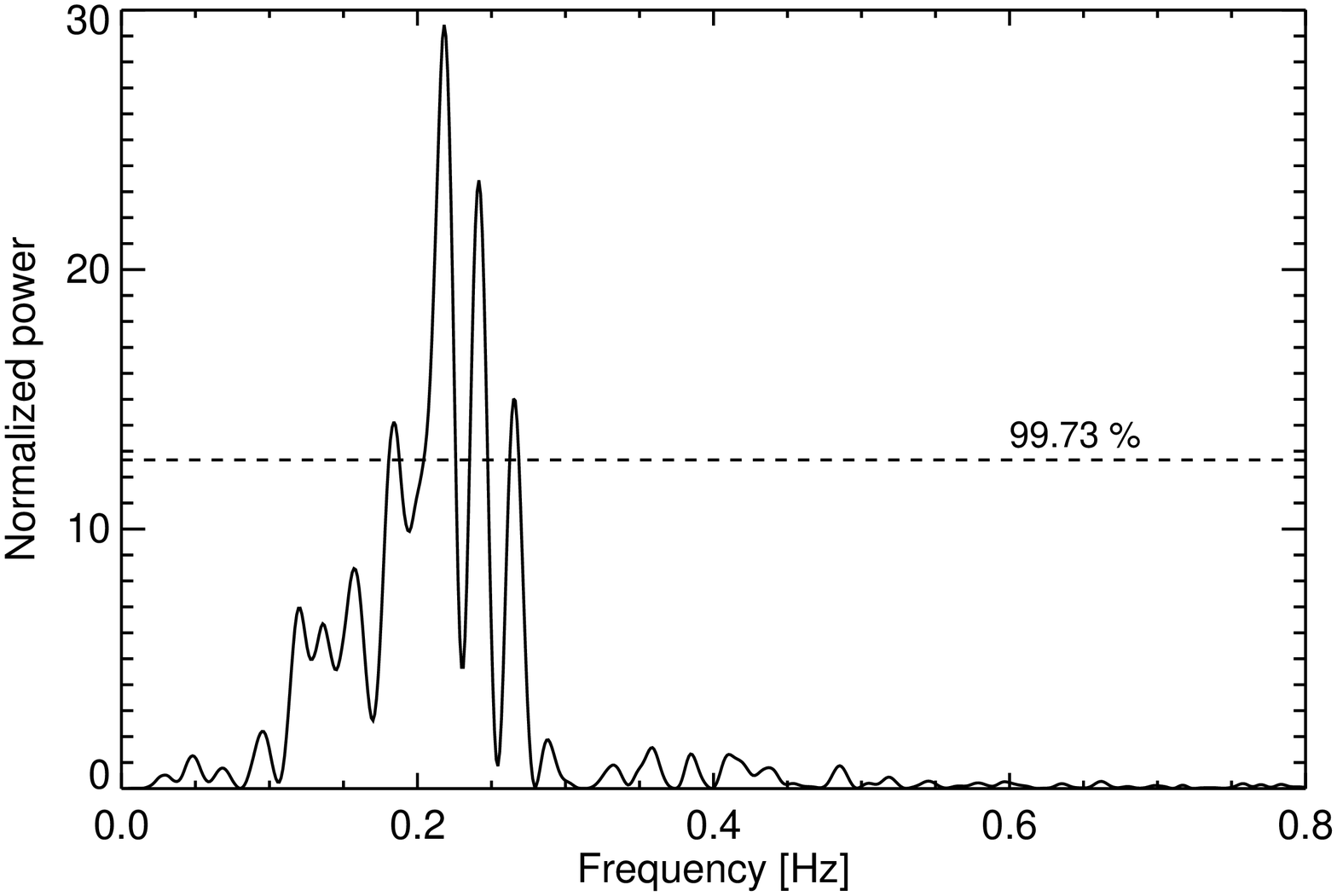}
\includegraphics[angle=0,width=0.5\textwidth]{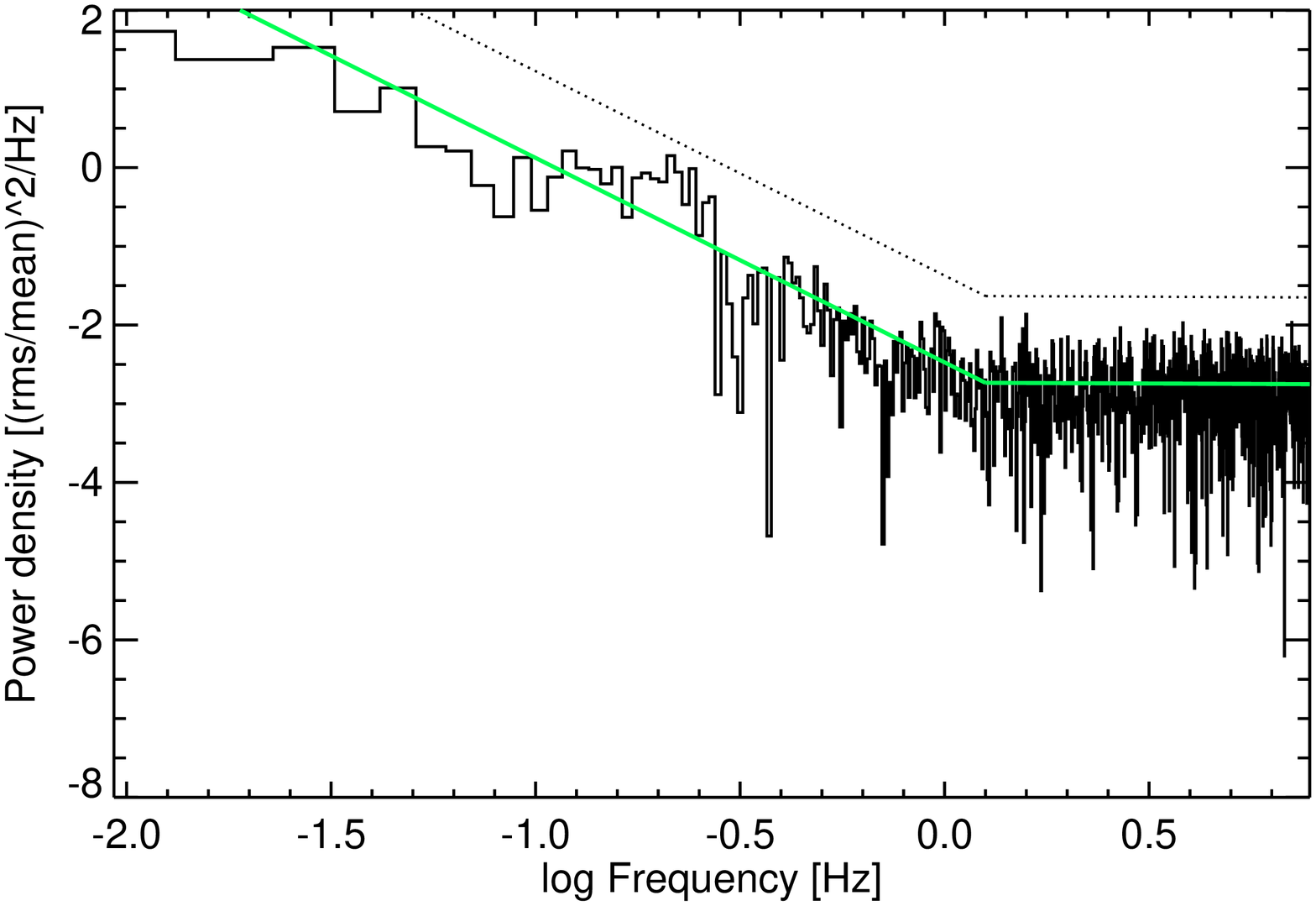}
\caption{\textit{Upper panel:} Same as Fig.\ref{fig:110224312_sfl} for the solar flare observed on March 15, 2011.} \textit{Middle panel:} Same as Fig.\ref{fig:110224312_sfl}. \textit{Lower panel:} Same as Fig.\ref{fig:110224312_sfl}. 
\label{fig:110315015_sfl}
\end{figure}

For the analysis we use again CTIME data of detectors NaI 0 through NaI 5 in an energy range covering 50~keV to 1~MeV with a time resolution of 0.256~s.
The total duration of the solar flare in this energy range is approximately 50~s (see Fig.\ref{fig:110315015_sfl}). For the Lomb-Scargle periodogram analysis (see middle panel of Fig.\ref{fig:110315015_sfl}) the light curve was detrended with a simple moving average of 5~s.
Repeating the PSD analysis on the original data set between ($T_0-30$)~s and ($T_0+100$)~s according to \citet{vaughan05} we did not find any significant QPP.

\subsection{Solar flare on March 14, 2011 at 19:50:17.3 UT}
\begin{figure}
\centering
\includegraphics[angle=0,width=0.5\textwidth]{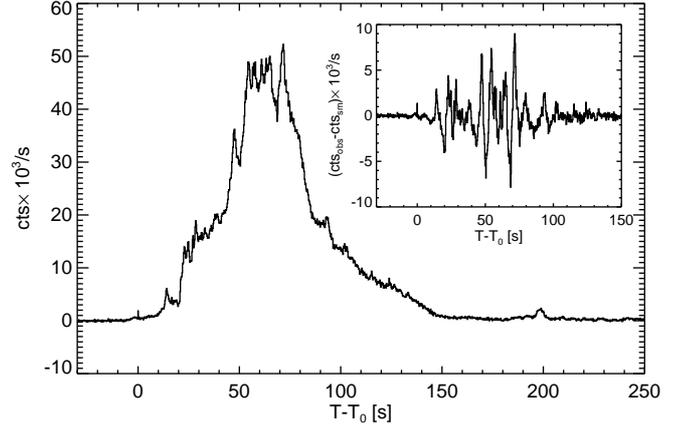}
\includegraphics[angle=0,width=0.5\textwidth]{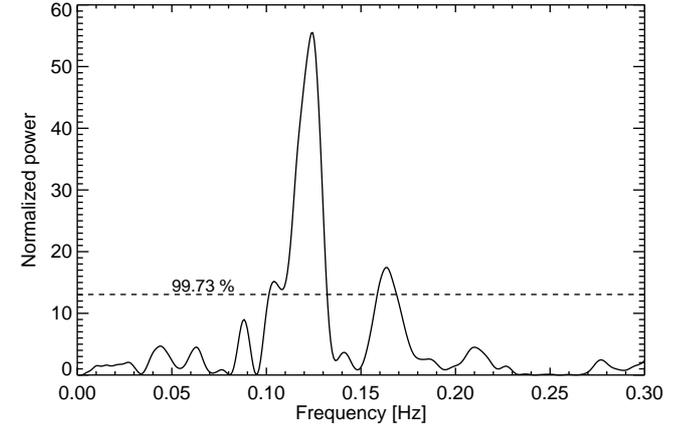}
\includegraphics[angle=0,width=0.5\textwidth]{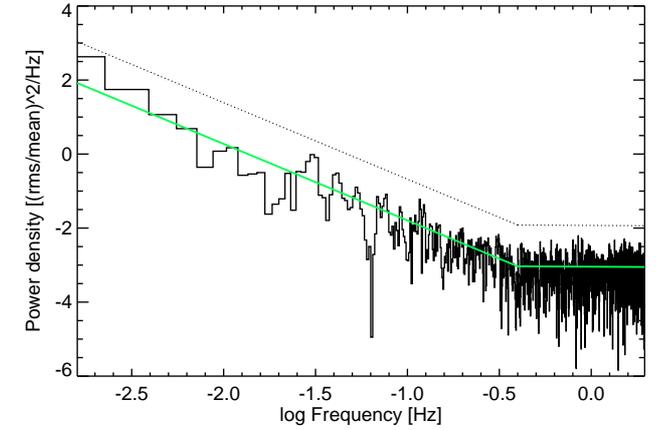}
\caption{\textit{Upper panel:} Same as Fig.\ref{fig:110224312_sfl} for the solar flare observed on March 14, 2011. \textit{Middle panel:} Same as Fig.\ref{fig:110224312_sfl}. \textit{Lower panel:} Same as Fig.\ref{fig:110224312_sfl}. }
\label{fig:110314827_sfl}
\end{figure}

For the analysis we use the same data type, energy range and time resolution as for solar flare \#4 in Sect.\ref{sec:sfl4}.
The total duration of the solar flare is approximately 150~s (see Fig.\ref{fig:110314827_sfl}).
Contrary to the a Lomb-Scargle periodogram which finds a significant periodicity in the detrended light curve (smoothing length of 10~s), the PSD, determined in the time interval between ($T_0-25$)~s and ($T_0+150$)~s and done according to \citet{vaughan05}, does not find any significant QPP.

\section{Summary \& Conclusions}
\label{sec:conc}
We have analyzed light curves of five solar flares observed by both \gbm and \rhessi.
We tested the data for the presence and significance of QPPs accounting for the overall shape of the PSD by 
applying the method introduced by \citet{vaughan05}. First of all, this technique was validated and tested by applying it on raw data of the \rhessi satellite. Any \rhessi light curve has an inherent period caused by the rotation of the space craft around its axis. With the method adopted here we successfully retrieve this well known 4~s period. However, we were not able to confirm the previously reported QPP of 40~s in the very same solar flare \citep{nak10}. An additional check was performed by applying the method to SPI-ACS data of the giant flare of the well known SGR~1806-20. This magnetar has a rotation period of $7.56$~s which, together with several harmonics, could be recovered unambiguously with the procedure presented here. These two tests gave us confidence that the method is appropriate to test the significance of QPPs in red-noise dominated solar flare time series.

The routine was then applied to four solar flares observed by \gbm. Although all of these solar flares showed very promising quasi-periodic features in their detrended light curves at first none of these were significant ($>$ 3~$\sigma$). Previous authors \citep[e.g.][]{inglis08,inglis09,zim10,nak10} who claim a significant QPP detection, applied a standard Lomb-Scargle analysis to detrended solar flare light curves. Our investigation of this method suggests the power in the low frequency range is being artificially suppressed, which can lead to misleading values for the significances of features in the PSD. A periodogram analysis should always be performed on the raw and undetrended light curve as was done here.

We stress once more that not only solar flares but many astrophysical sources (X-ray binaries, Seyfert galaxies, GRBs) suffer from steep power spectra in the 
low-frequency range. Such power spectra make a periodogram analysis not trivial, the interpretation of a peak in a PSD more complex and an estimation of its significance an important issue. Again we emphasize that red-noise is an intrinsic source property. In other words, having shown that the variations in the corresponding solar flares
 are not quasi-periodic at the 3$\sigma$ level does not
 mean that these variations are not real. The corresponding flux changes
 are sometimes dramatic, reaching a factor of a few within a few 
tens of seconds. Also, these variations occur in phase at different
 X-ray to gamma-ray energies, and other flares have been observed to
 also occur in phase with microwave radio emission \citep[e.g.][]{nak10,fou10}.

 While we think that it is not necessary to invoke oscillatory regimes
 of plasma instabilities \citep{nakingl10, rez11}, we note that the 
 physics of these variations is certainly interesting and worth further 
 studies.

\bibliographystyle{aa}
\bibliography{refs.bib}

\end{document}